\documentclass[eqsecnum,showpacs,amsmath,amssymb,superscriptaddress,nofootinbib,preprintnumbers]{revtex4-2}

\usepackage{graphicx}
\usepackage{color}

\begin{document}

\title{Exact one-loop QED actions in global (A)dS$_2$}

\author{Chiang-Mei Chen}
\email{cmchen@phy.ncu.edu.tw}
\affiliation{Department of Physics, National Central University, Zhongli, Taoyuan 320317, Taiwan}
\affiliation{Center for High Energy and High Field Physics (CHiP), National Central University, Zhongli, Taoyuan 320317, Taiwan}

\author{Sang Pyo Kim} \email{sangkim@kunsan.ac.kr}
\affiliation{Center for Relativistic Laser Science, Institute for Basic Science, Gwangju 61005, Korea}
\affiliation{ICRANet, 65122 Piazza della Repubblica, 10, Pescara, Italy}
\affiliation{School of Advanced Science and Technology, Kunsan National University, Kunsan 54150, Korea}

\author{Cristian Andres Rivera Medina} \email{cristianrivera@phy.ncu.edu.tw}
\affiliation{Department of Physics, National Central University, Zhongli, Taoyuan 320317, Taiwan}

\date{\today}

\begin{abstract}
Using the in–out formalism, we derive the exact one-loop QED effective actions for spinor field in a uniform electric field in two-dimensional global (anti-)de Sitter (A)dS$_2$ spacetime. The one-loop effective action probed by a scalar or spinor field is determined by the scattering matrix relating the out-vacuum to the in-vacuum, which is in turn fixed by the Bogoliubov coefficients of the corresponding Klein–Gordon or Dirac equation in the presence of both a gauge field and curved spacetime. Remarkably, the vacuum persistence amplitude---twice the imaginary part of the one-loop effective action---is related, via the Bogoliubov relations, to the mean number of particle–antiparticle pairs spontaneously produced by the background fields. The Bogoliubov coefficients or mean number of pair-production for charged scalar and spinor fields in global (A)dS$_2$ lead to QED effective actions expressed in terms of both proper-time integrals and Hurwitz zeta functions. These effective actions reveal a strong interplay between the electric field and spacetime curvature and correctly reproduce the limiting cases of pure (A)dS$_2$ spacetime and a uniform electric field in Minkowski space. We further discuss the physical implications and the invariant nature of the resulting QED effective actions in (A)dS$_2$.
\end{abstract}

\maketitle

\section{Introduction} \label{sec:intro}

Pairs of particles and antiparticles can be spontaneously produced in backgrounds of gauge fields and curved spacetimes. Schwinger pair production induced by strong electric fields~\cite{Schwinger:1951nm}, particle production in expanding universes~\cite{Parker:1968mv}, and Hawking radiation from black holes~\cite{Hawking:1975vcx} are the most well-known phenomena predicted by nonperturbative quantum field theory. In the in-out formalism
these backgrounds cause the out-vacuum properly defined in an asymptotic out-region to differ from the in-vacuum defined in an initial in-region. In fact, the out-vacuum can be expressed as a superposition of entangled multi-particle and antiparticle states built upon the in-vacuum: bosons contain arbitrarily many pairs, whereas fermions can have only one-pair or no-pair due to Pauli blocking. Moreover, the one-loop effective action for scalar or spinor fields is determined by the scattering matrix connecting the out- and in-vacua. The decay of the in-vacuum through spontaneous pair production then establishes a direct relation between the vacuum persistence amplitude and the mean number of pairs~\cite{DeWitt:1975ys, dewitt2003global}.

Charged black holes provide a theoretical arena for semiclassical quantum gravity, where Hawking radiation and the Schwinger effect jointly contribute to the emission of charged particles from the horizon~\cite{Gibbons:1975kk, Khriplovich:1999gm}.
In recent years, particular attention has been devoted to the Schwinger effect in (near-)extremal charged black holes, for which the Hawking temperature vanishes (or becomes negligibly small). In this regime, Schwinger pair production serves as the primary mechanism for charge emission and the discharging of these black holes~\cite{Kim:2008xv, Chen:2012zn, Chen:2020mqs, Cai:2020trh, Chen:2023swn, Chen:2024ctf, Chen:2025lnk, Betzios:2025sct, Brown:2024ajk, Mohan:2024rtn, Montero:2019ekk, Rakic:2025svg, Wang:2026oki, Lin:2024jug, Coviello:2025slv, Emparan:2025sao, Emparan:2025qqf}. Remarkably, the near-horizon geometry of (near-)extremal charged black holes exhibits enhanced symmetry of AdS$_2 \times S^2$ for extremal Reissner-Nordst\"{o}m black holes and dS$_2 \times S^2$ for charged Nariai black holes~\cite{Kunduri:2013gce}. This symmetry enhancement allows the Klein-Gordon and Dirac equations minimally coupled to a gauge field to be solved analytically in terms of special functions. Furthermore, spacetime curvature can either suppress or enhance Schwinger emission, depending on whether AdS$_2$ or dS$_2$. The Schwinger emission from near-extremal charged Nariai black holes exhibits bosonic amplification~\cite{Chen:2024ctf} but neither fermionic superradiance nor amplification~\cite{Chen:2025lnk} at one-loop. It is thus physically interesting to find one-loop effective actions for the charged scalar and spinor field in the (A)dS$_2$ spacetime, and to investigate the strong interaction or coupling between the electric fields and spacetime curvature.

When the scalar curvature and the Maxwell field become comparable to each other, their strong coupling raises important questions regarding the structure of the one-loop effective action. In particular, a consistent one-loop effective action should correctly reproduce the vacuum persistence amplitude, twice its imaginary part, and the Schwinger emission formula associated with spontaneous pair production by backgrounds. Finding one-loop effective actions in (A)dS spacetime and/or a uniform electric field has long been a theoretical challenge. Previous studies have addressed the effective action in pure (A)dS spacetime~\cite{Das:2006wg, Kim:2010cb, Akhmedov:2019esv, Akhmedov:2024axn, Jiang:2020evx, Zhou:2025jwm}, and scalar QED effective action in a uniform electric field in (A)dS~\cite{Cai:2014qba, Kim:2015azt}. In particular, the scalar QED action in (A)dS$_2$ derived within the in-out formalism~\cite{Cai:2014qba} reveals the consistency condition that relates the vacuum persistence to the mean number. Furthermore, a recent analysis~\cite{Zhou:2026elr} argues that the discrepancy between the Green's function and Bogoliubov methods arises because the former omits endpoint data, whereas the latter retains these contributions and thereby provides a complete description of the vacuum persistence probability.

The main purpose of this paper is to employ the in-out formalism to derive exact one-loop effective QED actions for spinor field in a uniform electric field in the global coordinates of (A)dS$_2$. A distinctive feature of the global geometry of the dS spacetime is that particle production occurs only in even dimensions, whereas it is prohibited in odd dimensions, regardless of scalar~\cite{Mottola:1984ar, Bousso:2001mw} or spinor~\cite{Jiang:2020evx}.
This property is closely related to the supersymmetry of the probing field in the global dS spacetime~\cite{Cooper:1994eh}. Recently, two of the authors (C.M.P. and S.P.K.) have studied Schwinger emission induced by a uniform electric field in the global (A)dS$_2$ and uncovered a reciprocal relation between the mean numbers of pairs produced in dS$_2$ and AdS$_2$ spacetimes~\cite{Chen:2025xrv}, all of which correspond to the vacuum persistence amplitude (imaginary part of effective action). These findings provide strong motivation for the present work that studies one-loop QED actions in global (A)dS$_2$.

There are two possible approaches to determine the one-loop QED actions within the in-out formalism. In the first approach, we should construct analytical functions that reproduce the correct imaginary parts of the effective action, express the QED actions in terms of proper-time integrals, and then rewrite them using Hurwitz zeta-functions. In the second approach, we directly evaluate the scattering matrix connecting the in- and out-vacua, which is given by the Bogoliubov coefficients. Remarkably, these coefficients in most solvable models take the form of ratios of gamma-functions with complex arguments that allow the QED actions to be expressed in the proper-time integral or the Hurwitz zeta-functions. In this paper, we apply the first method to find QED actions for spinor field in a uniform electric field in the global (A)dS$_2$. This method is more general, since various methods can directly give the mean number and thereby the vacuum persistence amplitude. The resulting QED actions are functionals of both the Maxwell field and the spacetime curvature, depending explicitly on the ratios of the electric field to the spacetime curvature and of the mass square to the curvature\footnote{It was argued for the scalar QED pair-production rate in (A)dS$_2$ planar coordinates~\cite{Pioline:2005pf} that the result depending only on invariant quantities is due to the highly symmetric configuration of a constant electric field in a maximally symmetric spacetime in 2D. This behavior was also suggested by the explicit dependence of the Bogoliubov coefficients on such invariants (up to a phase) found in~\cite{Chen:2025xrv} using global coordinates.}. They thus reveal a strong interaction between Maxwell theory and gravity at the one-loop. In contrast to results obtained from perturbative or derivative expansions~\cite{Drummond:1979pp, Davila:2009vt}, our effective actions possess an imaginary part that arises from the pole structure of the proper-time integral and that satisfies the consistency relation with the mean numbers or the so-called vacuum persistence amplitude.

The paper is organized as follows. In Sec.~\ref{sec2}, we review the in–out formalism, extract the one-loop effective action from the scattering matrix connecting the in- and out-vacua, and elucidate the relation between the imaginary part of the effective action and spontaneous pair production. In Sec.~\ref{sec3}, we apply Cauchy’s theorem to derive the proper-time representation of the one-loop effective action, which uniquely determines---up to entire functions---the analytic structure responsible for the desired imaginary part. We further obtain an alternative closed-form expression in terms of the Hurwitz zeta function by dimensionally regularizing the divergent proper-time integral. In Secs.~\ref{sec4} and~\ref{sec5}, we derive the QED effective action for a spinor field in both the proper-time and Hurwitz zeta-function representations. Using the proper-time form, we perform a perturbative expansion in terms of characteristic dimensionless parameters, namely the ratios of the Maxwell field strength to the spacetime curvature and of the mass squared to the curvature. This expansion correctly reproduces the weak-field limit of the Heisenberg–Euler and Schwinger QED actions in flat Minkowski spacetime. In the opposite limit of a vanishing electric field, after accounting for the density of states, the effective action reduces to the one-loop action in de Sitter space that is expressed as a curvature expansion beginning at order $R^2$. In Sec.~\ref{sec6}, we discuss the connection of QED actions between dS$_2$ and AdS$_2$. Finally, we summarize our results and present concluding remarks in Sec.~\ref{sec7}.

\section{One-Loop Effective Action in In-Out Formalism} \label{sec2}
In the in–out formalism~\cite{DeWitt:1975ys, dewitt2003global}, the one-loop effective action can be obtained from the scattering matrix of a probe field in a D-dimensional spacetime and/or a background gauge field, evaluated between the in- and out-vacua that are properly defined in the asymptotic in- and out-regions (in units of $\hbar = c = 1$)
\begin{eqnarray} \label{W_complex}
\mathrm{e}^{i W} = \exp\left( i \int d^Dx \sqrt{-g} \, L_\mathrm{eff}  \right) = \langle 0, \mathrm{out} | 0, \mathrm{in} \rangle.
\end{eqnarray}
The $W$ is complex in general but becomes real when the in-vacuum evolves trivially to the out-vacuum modulo a pure phase. In fact, twice the imaginary part of $W$, known as the vacuum persistence amplitude, is directly related to the mean number of spontaneously produced particle–antiparticle pairs, as will be shown below. Explicitly, it is expressed as
\begin{eqnarray} \label{vac_per_rel}
| \langle 0, \mathrm{out} | 0, \mathrm{in} \rangle |^2 = \mathrm{e}^{-2 \mathrm{Im} W}, \qquad 2 \mathrm{Im} W = \pm \sum_k \ln(1 \pm {\cal N}_k),
\end{eqnarray}
where ${\cal N}_k$ is the mean number of pairs for each mode $k$. Here and hereafter, the upper (lower) sign corresponds to the scalar (spinor) case. The $2 \mathrm{Im} W$ can be interpreted as the vacuum pressure of the ideal boson or fermion gas~\cite{Lebedev:1984mei}.

If one can construct an analytic complex function $W$ whose imaginary part coincides with $\mathrm{Im} W$, then its real part necessarily gives rise to the corresponding one-loop effective action~\cite{Kim:2016nyz}. This correspondence between the real and imaginary parts of $W$ should not be confused with the optical theorem of quantum field theory~\cite{peskin2018introduction}. Recently the dispersion relation of complex effective action by Heisenberg-Euler and Schwinger has been shown by expressing the imaginary and real parts in terms of quantum dilogarithm~\cite{Dunne:2025cyo}.
A caveat is that when the background does not spontaneously produce pairs of particles, the effective action becomes real. For instance, a charge in a constant magnetic field has Landau levels, and the one-loop effective action does not have the imaginary part~\cite{Schwinger:1951nm}. However, in the in-out formalism, the inverse scattering matrix directly gives the one-loop QED action that is real~\cite{Kim:2011cx}. Another example is the one-loop effective action of AdS space that does not have any pole for particle production~\cite{Das:2006wg}.

In this paper, we employ the in-out formalism to derive the one-loop effective actions, rather than other methods. This formalism relies on Bogoliubov transformations between the particle and antiparticle operators defined in two distinct asymptotic regions. When the in- and out-vacua are properly defined and the corresponding Fock spaces of particles and antiparticles are constructed, the out-vacuum is related to the in-vacuum through a Bogoliubov transformation,
\begin{equation}
\hat{a}_{k, \mathrm{out}} = \alpha_k \hat{a}_{k, \mathrm{in}} + \beta_k^* \hat{b}^\dag_{k, \mathrm{in}}, \qquad \hat{b}_{k, \mathrm{out}} = \alpha_k \hat{b}_{k, \mathrm{in}} \pm \beta_k^* \hat{a}^\dag_{k, \mathrm{in}}.
\end{equation}
Here, the in-vacuum (out-vacuum) is a state that contains neither particle nor antiparticle: $\hat{a}_{k, \mathrm{in}}| 0; \mathrm{in} \rangle = \hat{b}_{k, \mathrm{in}}| 0; \mathrm{in} \rangle = 0$ ($\hat{a}_{k, \mathrm{out}}| 0; \mathrm{out} \rangle = \hat{b}_{k, \mathrm{out}}| 0; \mathrm{out} \rangle = 0$). The mean number of pairs spontaneously produced is given by
\begin{eqnarray}
{\cal N}_k = \langle 0; \mathrm{in} \vert \hat{a}^{\dagger}_{k, \mathrm{out}} \hat{a}_{k, \mathrm{out}} \vert 0; \mathrm{in} \rangle = \langle 0; \mathrm{in} \vert \hat{b}^{\dagger}_{k, \mathrm{out}} \hat{b}_{k, \mathrm{out}} \vert 0; \mathrm{in} \rangle.
\end{eqnarray}

The annihilation and creation operators for a scalar (boson) satisfy the commutators for particle and antiparticle, respectively,
\begin{eqnarray}
[\hat{a}_{k', \mathrm{in}}, \hat{a}^{\dagger}_{k, \mathrm{in}}] = \delta_{k', k}, \qquad [\hat{b}_{k', \mathrm{in}}, \hat{b}^{\dagger}_{k, \mathrm{in}}] = \delta_{k', k},
\end{eqnarray}
and have the group structure $SU(1,1)$~\cite{perelomov1972boson}. Similarly, the commutators hold for the operators in the out-vacuum. Then, the commutators lead to the Bogoliubov relation
\begin{eqnarray}
| \alpha_k |^2 - | \beta_k |^2 = 1.
\end{eqnarray}
The out-vacuum can be expressed as a superposition of entangled particle–antiparticle states~\cite{Kim:2008yt, Ebadi:2014ufa},
\begin{eqnarray}
| 0; \mathrm{out} \rangle = \prod_k \frac1{\alpha_k} \sum_{n_k} \left( - \frac{\beta_k^*}{\alpha_k} \right)^{n_k} | n_k, \bar{n}_k; \mathrm{in} \rangle,
\end{eqnarray}
where the equality of particle and antiparticle occupation numbers follows from the conservation of charge and transverse momentum.
Hence, we can straightforwardly show that
\begin{eqnarray} \label{sc_ImW1}
\langle 0, \mathrm{out} | 0, \mathrm{in} \rangle = \prod_k (\alpha_k^*)^{-1}, \qquad | \langle 0, \mathrm{out} | 0, \mathrm{in} \rangle |^2 = \prod_k |\alpha_k|^{-2},
\end{eqnarray}
and therefrom,
\begin{eqnarray} \label{sc_ImW}
\mathrm{Im} W  = \frac{1}{2} \sum_k \ln |\alpha_k|^2 = \frac{1}{2} \sum_k \ln(1 + {\cal N}_k).
\end{eqnarray}

In the case of a spinor field, the annihilation and creation operators satisfy the anticommutators
\begin{eqnarray}
\{ \hat{a}_{k', \mathrm{in}}, \hat{a}^{\dagger}_{k, \mathrm{in}} \} = \delta_{k',k}, \qquad \{ \hat{b}_{k', \mathrm{in}}, \hat{b}^{\dagger}_{k, \mathrm{in}} \} = \delta_{k',k}
\end{eqnarray}
and give the Bogoliubov relation:
\begin{eqnarray}
| \alpha_k |^2 + | \beta_k |^2 = 1.
\end{eqnarray}
The fermionic operators have the group structure $SU(2)$~\cite{perelomov1972fermion}.
The Pauli exclusion principle restricts the occupation number of each mode and leads to the out-vacuum
\begin{eqnarray}
| 0; \mathrm{out} \rangle = \prod_k \Bigl( \alpha_k | 0_k, \bar{0}_k; \mathrm{in} \rangle - \beta_k^* | 1_k, \bar{1}_k; \mathrm{in} \rangle \Bigr).
\end{eqnarray}
The corresponding vacuum-to-vacuum amplitude is then
\begin{eqnarray} \label{sp_ImW1}
\langle 0, \mathrm{out} | 0, \mathrm{in} \rangle = \prod_k (\alpha_k^*), \qquad | \langle 0, \mathrm{out} | 0, \mathrm{in} \rangle |^2 = \prod_k |\alpha_k|^{2},
\end{eqnarray}
and therefore the imaginary part of the one-loop action is
\begin{eqnarray} \label{sp_ImW}
\mathrm{Im} W  = - \frac{1}{2} \sum_k \ln |\alpha_k|^2 = - \frac{1}{2} \sum_k \ln (1 - {\cal N}_k).
\end{eqnarray}

Thus in the in-out formalism, the vacuum persistence relation~\eqref{vac_per_rel} is a consistent condition that should be held, and we will use the relation to find the one-loop effective action from the mean number of spontaneously produced pairs. In many physical phenomena, such as Hawking radiation or Schwinger pair production by electric fields, various methods have been introduced to calculate the mean number. In the next sections, we will make use of~\eqref{vac_per_rel} to find the effective actions for fermions.

It should be noted that the difference between our approach and the direct treatment, in which $W = \pm i \ln \alpha^*$, is that the contribution of a phase is absent in our formulation. If the missing phase is reinstated by replacement $\alpha \to \alpha \exp(i W_{\rm ML})$, the effective action is shifted only in its real part, $W \to W \pm W_{\rm ML}$. As will be explicitly shown in Secs.~\ref{sec4} and~\ref{sec5}, the contributions from $W_{\rm ML}$ that depend on an angular quantum number appear exclusively as a phase $W_{\rm ML}$, modulo a magnitude independent of the angular number. Thus, the angular dependence does not appear in the imaginary part of the action. Mathematically, the Mittag-Leffler theorem states that the phase gives an entire function $W_{\rm ML}$.

In contrast to approaches based on mode decompositions in terms of quantum numbers, which typically yield a mode-by-mode effective action, recovering the full effective action requires an additional summation over all modes, thereby introducing another step of regularization~\cite{Jiang:2020evx, Pioline:2005pf, Castellano:2026ojb}. Our method avoids these technical complications by working directly within the Schwinger proper-time representation~\cite{Schwinger:1951nm}, where both the real and imaginary parts of the effective action in (A)dS$_2$ are obtained in a manifestly invariant framework.

\section{Proper-time Integral Representation} \label{sec3}

In the in-out formalism two promising approaches have been advanced to find the one-loop effective action: either from the Bogoliubov coefficients or mean numbers. The first approach is to compute the complex effective action directly from Eq.~\eqref{W_complex} and then extract its real part. The second approach is to compute the imaginary part of the effective action from the mean numbers as in Eq.~\eqref{vac_per_rel} and subsequently reconstruct the corresponding real part via analytic continuation of the complex action modulo entire functions via the Mittag-Leffler theorem~\cite{Markushevich1985}. In the latter approach, the evaluation of Eq.~\eqref{vac_per_rel} involves a relatively simple expression in terms of real functions, typically of the form $\ln(1 \pm \mathrm{e}^{-S})$ with $S$ denoting an instanton action for pair production. By constructing an appropriate contour integral, one can then recover both the complex and real parts of the effective action.
The second approach here shares a similar concept of Borel summation of the imaginary part, in which the simple poles of a proper-time integral give the imaginary part and the real part is the effective action (for instance, see~\cite{Dunne:1999vd} and references therein.)

\begin{figure}[ht]
\centerline{\includegraphics[scale=0.7, angle=0]{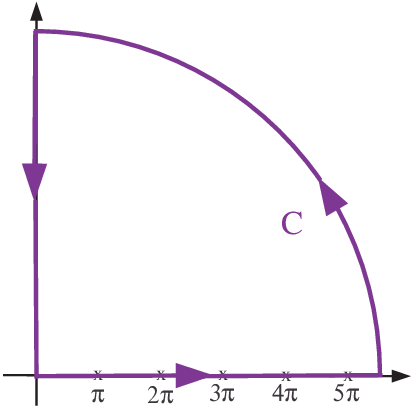}}
\caption{The contour for the integrations~\eqref{eq_ConI} and~\eqref{eq_ConII}.}
\label{fig_contour}
\end{figure}

In formulating the second approach, an important mathematical issue is to find a contour integration for the function
\begin{equation}
\ln\left( 1 + \mathrm{e}^{-S} \right) = \mathrm{e}^{-S} - \frac12 \mathrm{e}^{-2S} + \frac13 \mathrm{e}^{-3S} - \frac14 \mathrm{e}^{-4S} + \cdots,
\end{equation}
as a sum of residues in the Cauchy's residue theorem. The well-known pole expansion of $\csc z$
\begin{equation}
\csc z = \frac1{\sin z} = \frac1{z} - 2 z \left( \frac1{z^2 - \pi^2} - \frac1{z^2 - (2 \pi)^2} + \frac1{z^2 - (3 \pi)^2} - \cdots \right),
\end{equation}
leads to the construction of the following function, with real and positive parameter $S$,
\begin{equation} \label{csc_pole}
- \mathrm{e}^{-S z/\pi} \left( \frac1{\sin z} - \frac1{z} \right) \frac1{z},
\end{equation}
which contains simple poles at $z = n \pi$ with residues $(-1)^{n-1} \mathrm{e}^{-n S}/n \pi, n = 1, 2, 3, \cdots$. Note that the simple pole at $z = 0$ is removed in~\eqref{csc_pole}, which corresponds to excluding the origin of a proper-time integral, as will be shown below. A caveat arises from the Mittag-Leffler theorem, which states that~\eqref{csc_pole} holds up to entire functions in the complex plane~\cite{Markushevich1985}. Such entire functions will be neglected since they do not contribute to the renormalization of physical quantities.
Consequently, the contour integral, as shown in Fig.~\ref{fig_contour}, of this function gives the following relation
\begin{equation} \label{eq_ConI}
- \oint_C \mathrm{e}^{- S z/\pi} \left( \frac1{\sin z} - \frac1{z} \right) \frac{dz}{z} = i \ln\left( 1 + \mathrm{e}^{-S} \right).
\end{equation}
There are three parts in the contour integration: (i) along the positive real axis (real), (ii) along infinite arc (vanishing), and (iii) along imaginary axis (complex), so the above contour integration can be rewritten as
\begin{equation} \label{eq_ln+}
- i \int_0^\infty \mathrm{e}^{- i S y/\pi} \left( \frac1{\sinh y} - \frac1{y} \right) \frac{dy}{y} = P \int_0^\infty \mathrm{e}^{- S x/\pi} \left( \frac1{\sin x} - \frac1{x} \right) \frac{dx}{x} + i \ln\left( 1 + \mathrm{e}^{-S} \right).
\end{equation}
Here, the principal value integral on the right hand side is real and corresponds to the effective action.

Similarity, for the function $\ln(1 - \mathrm{e}^{-S})$, instead of $1/\sin z$, one should use $\cos z/\sin z$ to have non-alternating values of residues. Hence, the following integration along the same contour
\begin{equation} \label{eq_ConII}
- \oint_C \mathrm{e}^{- S z/\pi} \left( \frac{\cos z}{\sin z} - \frac1{z} \right) \frac{dz}{z} = i \ln\left( 1 - \mathrm{e}^{-S} \right),
\end{equation}
gives
\begin{equation} \label{eq_ln-}
- i \int_0^\infty \mathrm{e}^{- i S y/\pi} \left( \frac{\cosh y}{\sinh y} - \frac1{y} \right) \frac{dy}{y} = P \int_0^\infty \mathrm{e}^{- S x/\pi} \left( \frac{\cos x}{\sin x} - \frac1{x} \right) \frac{dx}{x} + i \ln\left( 1 - \mathrm{e}^{-S} \right).
\end{equation}

Interestingly, this approach automatically incorporates a specific regularization, namely the subtractions of $1/y$ and $1/x$ in the proper-time integrals, Eqs.~\eqref{eq_ln+} and~\eqref{eq_ln-}, which are known as the Schwinger subtraction-scheme in two dimensions~\cite{Schwinger:1951nm}.
However, in Minkowski spacetime of dimensions higher than two, the instanton action $S$ in a constant electric field has an additional factor whose transverse-momentum integral introduces strong divergences $1/y^n$ and $1/x^n$ ($n \geq 2$), which require more subtraction terms to make the integrals finite~\cite{Kim:2008yt}.

To illustrate these two approaches, let us consider a simple system of spinor pair production with the following Bogoliubov coefficients:
\begin{equation}
\alpha^* = \frac{\Gamma^2\left( 1/2 + i S/2\pi \right)}{\Gamma\left( 1 + i S/2\pi \right) \Gamma\left( i S/2 \pi \right)}, \qquad \beta = \frac{\Gamma^2\left( 1/2 + i S/2\pi \right)}{\pi}.
\end{equation}
These coefficients satisfy $|\alpha|^2 + |\beta|^2 = 1$, and the mean number of produced spinor pairs is
\begin{equation}
{\cal N} = |\beta|^2 = \frac{1}{\cosh^2(S/2)}.
\end{equation}
The imaginary part of the effective action is, therefore,
\begin{equation}
2 \mathrm{Im} W = - \ln\left( 1 - \mathcal{N} \right) = - \ln\left( \frac{\sinh^2(S/2)}{\cosh^2(S/2)} \right) = \ln\left( \frac{1 + \mathrm{e}^{-S}}{1 - \mathrm{e}^{-S}} \right)^2.
\end{equation}
The corresponding effective action can be obtained by subtracting Eq.~\eqref{eq_ln+} from Eq.~\eqref{eq_ln-}:
\begin{eqnarray} \label{sp-mod}
W_{(1)} &=& - i \int_0^\infty \mathrm{e}^{- i S y/\pi} \left( \frac{1}{\sinh y} - \frac{\cosh y}{\sinh y} \right) \frac{dy}{y}
\nonumber\\
&=& P \int_0^\infty \mathrm{e}^{- S x/\pi} \left( \frac{1}{\sin x} - \frac{\cos x}{\sin x} \right) \frac{dx}{x} + i \ln\left( \frac{1 + \mathrm{e}^{-S}}{1 - \mathrm{e}^{-S}} \right).
\end{eqnarray}

We can also obtain the effective action for spinors using an alternative approach, namely $W = -i \ln(\alpha^*)$ in Eq.~\eqref{sp_ImW1}. Employing the Gamma-function regularization scheme~\cite{Kim:2008yt}, we find
\begin{eqnarray} \label{ln-alp}
W_{(2)} = - i \ln (\alpha^*) &=& -i \Biggl[ 2 \ln \Gamma\left( \frac{1}{2} + i \frac{S}{2 \pi} \right) - \ln \Gamma\left( 1 + i \frac{S}{2 \pi} \right) - \ln \Gamma\left( i \frac{S}{2 \pi} \right) \Biggr]
\nonumber\\
&=& -i \int_{0}^{\infty} \frac{dy}{y} \mathrm{e}^{-i S y/2 \pi} \Bigl( \frac{2 \mathrm{e}^{-y/2}}{1 - \mathrm{e}^{-y}} - \frac{\mathrm{e}^{-y}}{1 - \mathrm{e}^{-y}} - \frac{1}{1 - \mathrm{e}^{-y}} - \cdots \Bigr),
\end{eqnarray}
where the ellipsis denotes subtraction terms required for renormalization in the Schwinger scheme. The regularized form of Eq.~\eqref{ln-alp} becomes
\begin{eqnarray}
W_{(2)} &=& i \int_{0}^{\infty} \frac{dy}{y} \mathrm{e}^{-i S y/2 \pi} \frac{(1 -  \mathrm{e}^{-y/2})^2}{(1 - \mathrm{e}^{-y/2}) (1 + \mathrm{e}^{-y/2})} = i \int_{0}^{\infty} \frac{dy}{y} \mathrm{e}^{-i S y/\pi} \frac{1 -  \mathrm{e}^{-y}}{1 + \mathrm{e}^{-y}}
\nonumber\\
&=& i \int_{0}^{\infty} \frac{dy}{y} \mathrm{e}^{-i S y/\pi} \frac{\sinh(y/2)}{\cosh(y/2)} = - i \int_0^\infty \mathrm{e}^{-i S y/\pi} \left( \frac{1}{\sinh y} - \frac{\cosh y}{\sinh y} \right) \frac{dy}{y}.
\end{eqnarray}
Two approaches give the same effective action $W_{(1)} = W_{(2)}$.

To obtain a more compact expression in terms of a finite set of special functions rather than integral representations, it is convenient to employ the Hurwitz zeta function and dimensional regularization~\cite{Elizalde:1994gf} to extract the finite parts of Eqs.~\eqref{eq_ln+} and~\eqref{eq_ln-}. Dimensional regularization was used for four dimensional QED actions by Heisenberg-Euler and Schwinger in a constant magnetic field and a constant electric field~\cite{Dittrich:1975au}. For our purpose, the integrals appearing in the first terms can be evaluated using standard formulas from Ref.~\cite{gradshteyn2014table}
\begin{eqnarray}
&& \int_0^\infty x^{\nu - 1} \mathrm{e}^{- \lambda x} \coth x \, dx = \Gamma(\nu) \left[ 2^{1 - \nu} \zeta(\nu, \lambda/2) - \lambda^{-\nu} \right],
\\
&& \int_0^\infty x^{\nu - 1} \mathrm{e}^{- \lambda x} \frac1{\sinh x} \, dx = 2^{1 - \nu} \Gamma(\nu) \zeta(\nu, \lambda/2 + 1/2).
\end{eqnarray}
Here, $\zeta(\nu, \lambda/2)$ is the Hurwitz zeta function.

To implement dimensional regularization, we first generalize the integral in Eq.~\eqref{eq_ln-} by introducing a regulator $n$ (setting $n = 0$ at the end of calculation) and defining $a = i S/2\pi$:
\begin{equation}
I = \int_0^\infty \mathrm{e}^{- 2 a y} \left( y^{n - 1} \coth y - y^{n - 2} \right) dy = \Gamma(n) \left[ 2^{1 - n} \zeta(n, a) - (2 a)^{-n} \right] - \Gamma(n - 1) (2 a)^{1 - n}.
\end{equation}
In the spirit of dimensional regularization, we set $n = 0 + \epsilon$ and, using the recursion formula of Gamma function, we obtain
\begin{equation}
I = \frac{\Gamma(\epsilon + 1)}{\epsilon} \left[ 2^{1 - \epsilon} \zeta(\epsilon, a) - (2 a)^{-\epsilon} \right] - \frac{\Gamma(\epsilon + 1)}{\epsilon (\epsilon - 1)} (2 a)^{1 - \epsilon},
\end{equation}
which diverges as $\epsilon \to 0$. Separating the divergent and finite parts,
\begin{equation}
I \simeq \frac{I_\mathrm{div}}{\epsilon} + I_\mathrm{reg}, \quad \to \quad I_\mathrm{reg} \simeq \frac{\epsilon I - I_\mathrm{div}}{\epsilon} \simeq \frac{d}{d\epsilon} (\epsilon I)\Big|_{\epsilon = 0},
\end{equation}
we obtain the regularized result
\begin{equation}
I_\mathrm{reg} = 2 \zeta'(0, a) + (1 - 2 a) \ln a + 2 a = 2 \ln\Gamma(a) - \ln(2 \pi) + (1 - 2 a) \ln a + 2 a.
\end{equation}

Similarly, for Eq.~\eqref{eq_ln+}, the integral can be expressed as
\begin{equation}
I = \frac{\Gamma(\epsilon + 1)}{\epsilon} 2^{1 - \epsilon} \zeta(\epsilon, a + 1/2) - \frac{\Gamma(\epsilon + 1)}{\epsilon (\epsilon - 1)} (2 a)^{1 - \epsilon},
\end{equation}
which gives the regularized expression
\begin{equation}
I_\mathrm{reg} = 2 \zeta'(0, a + 1/2) - 2 a \ln a + 2 a = 2 \ln\Gamma(a + 1/2) - \ln(2\pi) - 2 a \ln a + 2 a.
\end{equation}

\section{Effective Actions of Spinor QED in Global dS$_2$} \label{sec4}

For studying the Schwinger effect in global dS spacetime, the external constant electric field prevents both KG and Dirac equations from being separated, except for dS$_2$ with the scalar curvature $R = 2 H^2$
\begin{eqnarray} \label{dS2_metric}
ds^2 = - dt^2 + \frac{\cosh^2(H t)}{H^2} d\theta^2, \qquad -\infty < t < \infty, \quad 0 \le \theta \le 2 \pi.
\end{eqnarray}
So, we consider a constant electric field with the gauge potential for dS$_2$
\begin{eqnarray} \label{dS2_A}
A = - \frac{E \sinh(H t)}{H^2} d\theta \quad \Rightarrow \quad dA = - E \, \vartheta^t \wedge \vartheta^\theta,
\end{eqnarray}
where $\vartheta^t = dt$ and $\vartheta^\theta = \cosh(H t) d \theta/H$. Then the electric field points in the positive (clockwise) direction of $\theta$.

The Bogoliubov coefficients that describe spinor pair production in global dS$_2$ with a constant electric field were computed in~\cite{Chen:2025xrv}. The mean number of produced particles, modulo the spin multiplicity, is given by
\begin{equation} \label{spinor_ds}
{\cal N}^{\rm (sp)}_{\rm dS} = |\beta|^2 = \Bigl( \frac{\cosh \pi \kappa}{\cosh \pi \mu} \Bigr)^2.
\end{equation}
The corresponding vacuum persistence amplitude satisfies the Bogoliubov relations
\begin{eqnarray}\label{spinor_ds_alp}
|\alpha|^2 = \frac{\sinh (\pi \mu - \pi \kappa) \sinh (\pi \mu + \pi \kappa)}{\cosh^2\pi \mu}, \qquad |\alpha|^2 + |\beta|^2 = 1,
\end{eqnarray}
where the parameters $\kappa$ and $\mu$ are defined as
\begin{equation} \label{parameter}
\kappa = q E/H^2, \qquad \mu = \sqrt{\kappa^2 + m^2/H^2}.
\end{equation}
According to Eq.~\eqref{sp_ImW}, the imaginary part of the effective action can be written as a sum of three logarithmic terms
\begin{equation} \label{sp_vac_per}
\mathrm{Im} W^{\rm (sp)}_{\rm dS} = - \frac12 \ln\left( 1 - \mathrm{e}^{-2 \pi (\mu - \kappa)} \right) - \frac12 \ln\left( 1 - \mathrm{e}^{-2 \pi (\mu + \kappa)} \right) + \ln\left( 1 + \mathrm{e}^{-2 \pi \mu} \right).
\end{equation}
Using the relations~\eqref{eq_ln+} and~\eqref{eq_ln-}, the complex effective action is
\begin{eqnarray} \label{cW_dS}
W^{\rm (sp)}_{\rm dS} &=& \frac{i}{2} \int_0^\infty \left( \mathrm{e}^{- 2 i (\mu - \kappa) y} + \mathrm{e}^{- 2 i (\mu + \kappa) y} \right) \left( \frac{\cosh y}{\sinh y} - \frac1{y} \right) \frac{dy}{y}
\nonumber\\
&-& i \int_0^\infty \mathrm{e}^{- 2 i \mu y} \left( \frac1{\sinh y} - \frac1{y} \right) \frac{dy}{y}.
\end{eqnarray}
Thus, the real part of effective action in the proper-time integral that has the same imaginary part from simple poles follows from~\eqref{eq_ln+} and~\eqref{eq_ln-}:
\begin{eqnarray} \label{sp_vac_act}
\mathrm{Re} W^{\rm (sp)}_{\rm dS} &=& - \frac12 P \int_0^\infty \left( \mathrm{e}^{-2 (\mu - \kappa) x} + \mathrm{e}^{-2 (\mu + \kappa) x} \right) \left( \frac{\cos x}{\sin x} - \frac1{x} \right) \frac{dx}{x}
\nonumber\\
&+& P \int_0^\infty \mathrm{e}^{-2 \mu x} \left( \frac1{\sin x} - \frac1{x} \right) \frac{dx}{x}.
\end{eqnarray}
Note that the expression for $\mathrm{Re} W^{\rm (sp)}_{\rm dS}$ contains two noteworthy contributions. The first integral has the same functional form as the spinor QED effective action in a constant electric field in two-dimensional Minkowski spacetime under the identification $2 (\mu \mp \kappa) \leftrightarrow m^2/q E$, while the second integral matches the scalar QED effective action under the correspondence $2 \mu \leftrightarrow m^2/q E$~\cite{Kim:2008yt}. In fact, the Heisenberg-Euler and Schwinger QED actions in two dimensions have the vacuum persistence amplitude for the spinor (scalar) pair production: $2 \mathrm{Im} W_{\rm HES} = \mp \ln(1 \mp \mathrm{e}^{-\pi m^2/qE})$.
These two integrals together reproduce the correct vacuum persistence amplitude and the mean number of produced pairs for a spinor field in global dS$_2$ spacetime.

Since the exponential functions in $\mathrm{Re} W^{\rm (sp)}_{\rm dS}$~\eqref{sp_vac_act} decay sufficiently rapidly and play regulators of the integrals, the dominant contributions to the integrals arise from the region of small values of the integration variable. In this case, the integrations can be performed order by order using the following expansions~\cite{gradshteyn2014table}
\begin{equation} \label{eq_CS}
\frac{\cos x}{\sin x} - \frac1{x} = \sum_{n =1}^{\infty} \frac{(-1)^{n} 2^{2n} B_{2n}}{(2n)!} x^{2n-1} = -\frac13 x - \frac1{45} x^3 - \frac2{945} x^5 - \cdots,
\end{equation}
and
\begin{equation}
\frac{1}{\sin x} - \frac1{x} = \sum_{n =1}^{\infty} \frac{(-1)^{n+1} 2(2^{2n-1} -1 ) B_{2n}}{(2n)!} x^{2n-1} = \frac16 x + \frac7{360} x^3 + \frac{31}{15120} x^5 + \cdots,
\end{equation}
in which $B_{2n}$ are the Bernoulli numbers.
Similarly to the weak-field expansion of Schwinger effective action~\cite{Schwinger:1951nm}, we obtain the first few leading terms
\begin{eqnarray} \label{sp_exp}
\mathrm{Re} W^{\rm (sp)}_{\rm dS} &=& \frac{3 \mu^2 - \kappa^2}{12 \mu (\mu^2 - \kappa^2)} + \frac{15 \mu^6 + 3 \kappa^2 \mu^4 + 21 \kappa^4 \mu^2 - 7 \kappa^6}{1440 \mu^3 (\mu^2 - \kappa^2)^3}
\nonumber\\
&+& \frac{63 \mu^{10} + 165 \kappa^2 \mu^8 + 470 \kappa^4 \mu^6 - 310 \kappa^6 \mu^4 + 155 \kappa^8 \mu^2 - 31 \kappa^{10}}{20160 \mu^5 (\mu^2 - \kappa^2)^5} + \cdots.
\end{eqnarray}
Note that Eq.~\eqref{sp_exp} exhibits a nontrivial coupling between the electric field and the spacetime curvature, since Eq.~\eqref{parameter} involves both the ratio of the electric field to the curvature and the ratio of the mass squared to the curvature.

Moreover, we may consider two limiting cases: the weak-curvature limit and the weak-field limit. In the weak-curvature limit $H \to 0$, both $\kappa$ and $\mu$ diverge while their difference approaches $\mu - \kappa \to m^2/2 q E$. Consequently, only the first exponential term in~\eqref{cW_dS} contributes nontrivially to the effective action; the remaining terms yield only constant (field-independent) contributions. The effective action therefore reduces to
\begin{equation} \label{cW_Min}
W^{\rm (sp)}_{\rm Min} = \frac{i}2 \int_0^\infty \mathrm{e}^{-i m^2 y/q E} \left( \frac{\cosh y}{\sinh y} - \frac1{y} \right) \frac{dy}{y},
\end{equation}
which can be directly compared with the complex QED effective Lagrangian for a constant electric field in two-dimensional Minkowski spacetime~\cite{Kim:2008yt},
\begin{eqnarray}
{\cal L}^{\rm (sp)}_{\rm Min} = i \frac{qE}{4 \pi} \int_0^\infty  \mathrm{e}^{- i m^2 s/qE} \left( \frac{\cosh s}{\sinh s} - \frac{1}{s} \right) \frac{ds}{s}.
\end{eqnarray}
Here $q E/2 \pi$ is the density of states, which follows from $\mathcal{D} = \mu H^2/2 \pi$ in the limit $H \to 0$.\footnote{The Lagrangian density is $\mathcal{L} = \mathcal{D} W$.} Furthermore, the expansion of the real part of the one-loop action in~\eqref{sp_exp} becomes
\begin{equation}
\mathrm{Re} W^{\rm (sp)}_{\rm dS} \Big|_{H \to 0} = \frac{q E}{6 m^2} + \frac{q^3 E^3}{45 m^6} + \frac{8 q^5 E^5}{315 m^{10}} + \left( \frac1{6 q E} + \frac{q E}{60 m^4} + \frac{2 q^3 E^3}{63 m^8} \right) H^2 + \mathcal{O}(H^4).
\end{equation}
The vanishing-curvature limit $H = 0$ therefore correctly reproduces the standard one-loop QED action in Minkowski spacetime.

In contrast, in the weak-field limit $E \to 0$ the complex one-loop action~\eqref{cW_dS} reduces to a different form
\begin{eqnarray}
W^{\rm (sp)}_{\rm dS} \Big|_{E=0} = i \int_0^\infty \mathrm{e}^{- 2 i m y/H} \left( \frac{\cosh y}{\sinh y} - \frac{1}{\sinh y} \right) \frac{dy}{y}.
\end{eqnarray}
Using Eqs.~\eqref{eq_ln+} and~\eqref{eq_ln-}, we can extract the real and imaginary parts of the one-loop effective action in dS$_2$:
\begin{eqnarray} \label{eff_pure_dS}
W^{\rm (sp)}_{\rm dS} \Big|_{E=0} = - P \int_0^\infty \mathrm{e}^{- 2 m x/H} \left( \frac{\cos x}{\sin x} - \frac{1}{\sin x} \right) \frac{dx}{x} - i \ln \left( \frac{1 - \mathrm{e}^{-2 \pi m/H}}{ 1 + \mathrm{e}^{-2 \pi m/H}} \right).
\end{eqnarray}
Note that the effective action in pure dS space takes the form of a difference between scalar and spinor QED actions in a constant electric field~\cite{Kim:2008yt}, in the correspondence $H/2 \leftrightarrow qE/m$.
The vacuum persistence amplitude reduces to the standard result for spinor pair production in global coordinates,
\begin{eqnarray}
2 \mathrm{Im} W^{\rm (sp)}_{\rm dS} = - \ln (1- {\cal N}_{\rm dS}), \qquad {\cal N}_{\rm dS} = \frac{1}{\cosh^2 (\pi m/H)}.
\end{eqnarray}
In addition, the real part of the effective action admits the weak-field series expansion
\begin{equation}
\mathrm{Re} W^{\rm (sp)}_{\rm dS} \Big|_{E \to 0} = \frac{H}{4 m} + \frac{H^3}{96 m^3} + \frac{H^5}{320 m^5} + \left( \frac1{24 m^3 H} + \frac{17 H}{960 m^5} + \frac{43 H^3}{2688 m^7} \right) (q E)^2 + \mathcal{O}(E^4).
\end{equation}
The vanishing-field limit $E = 0$ therefore reproduces the one-loop effective action of pure global dS$_2$ spacetime.

\begin{figure}[t]
\centering
\setlength{\fboxrule}{1.2pt} 
\setlength{\fboxsep}{6pt} 

\begin{minipage}{0.48\textwidth}
\centering
\fbox{%
\begin{minipage}{\linewidth}
\centering
\includegraphics[width=\linewidth]{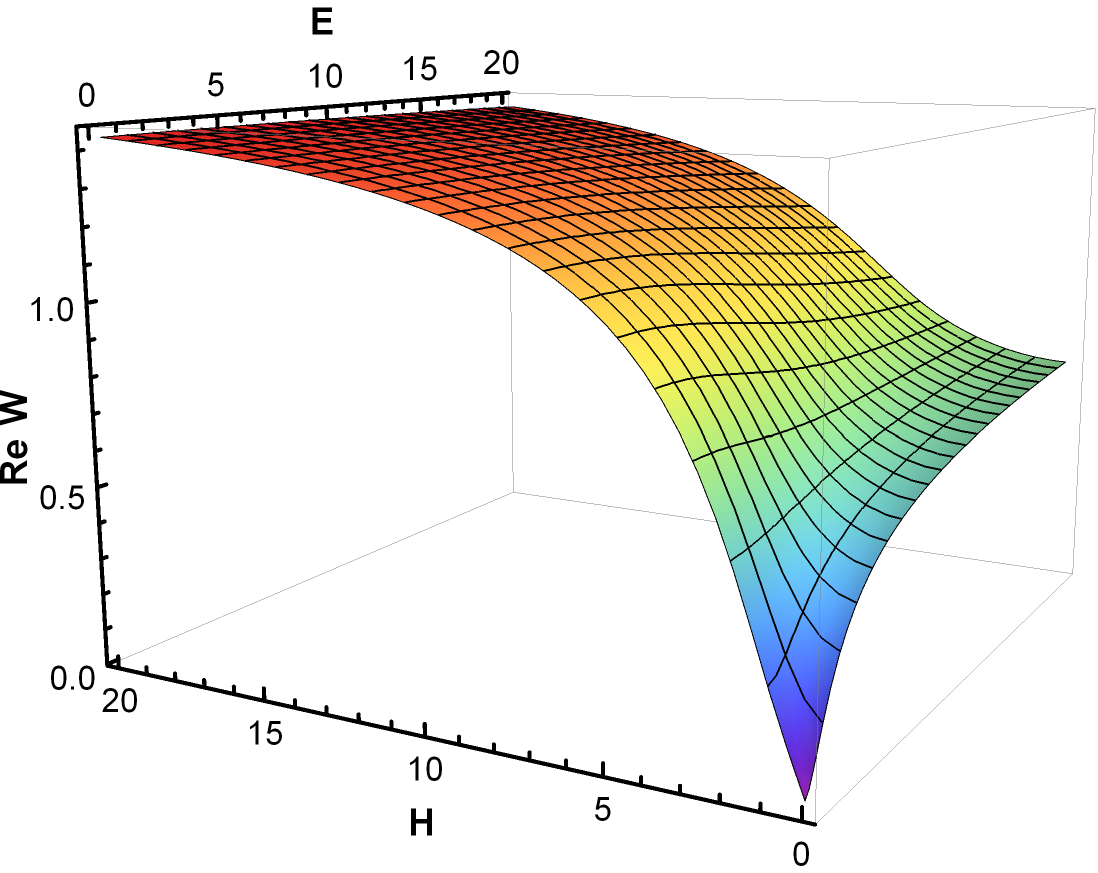}
\vspace{2mm}\textbf{(a)}\par
\end{minipage}}
\end{minipage}
\hfill
\begin{minipage}{0.48\textwidth}
\centering
\fbox{%
\begin{minipage}{\linewidth}
\centering
\includegraphics[width=\linewidth]{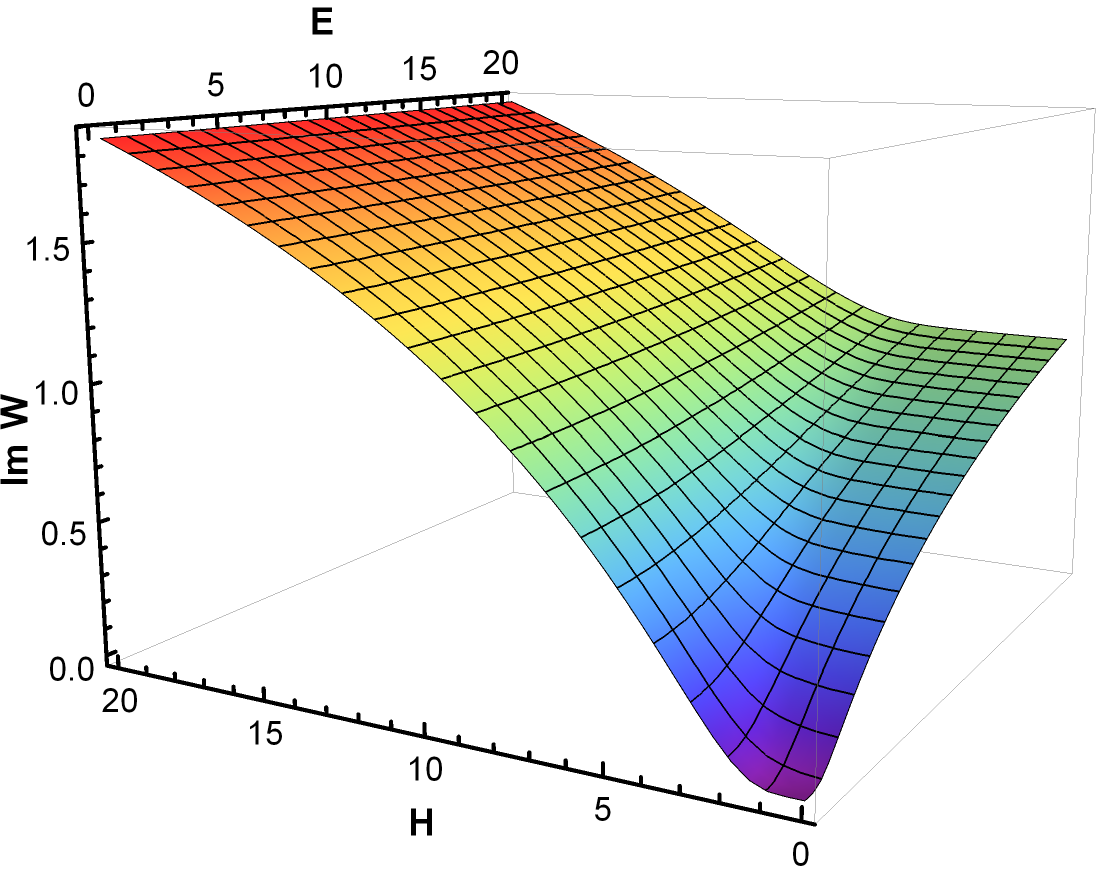}
\vspace{2mm}\textbf{(b)}\par
\end{minipage}}
\end{minipage}

\caption{The plots of real part (a) and imaginary part (b) of $\mathcal{W}^{\rm (sp)}_{\rm dS}$ against $E$ and $H$ for spinor QED in dS$_2$ in unit of $m = 1$ and $q = 1$. [Left panel] The one-loop effective action increases from $\mathrm{Re} \mathcal{W}^{\rm (sp)}_{\rm dS}(E=0, H=0) = 0$ as $E$ and $H$ increase. The increase is more significant for $H$ than $E$, but the action almost saturates for large $H$ regardless of $E$. The curvature effect ($R =2 H^2$) is more noticeable for weak field than for strong electric field.
[Right panel] The vacuum decays significantly as $H$ increases and gently as $E$ increases.   }
\label{fig:ReImSpinordS2}
\end{figure}

An alternative representation of the complex one-loop effective action is obtained by employing dimensional regularization, leading to an expression in terms of the Hurwitz zeta function
\begin{eqnarray}
\mathcal{W}^{\rm (sp)}_{\rm dS} &=& i \Bigl[ \zeta'(0, i \mu - i \kappa) + \zeta'(0, i \mu + i \kappa) - 2 \zeta'(0, 1/2 + i \mu) \Bigr]
\\
&+& i \Bigl[ (1/2 - i \mu + i \kappa) \ln(i \mu - i \kappa) + (1/2 - i \mu - i \kappa) \ln(i \mu + i \kappa) + 2 i \mu \ln(i \mu) \Bigr].
\nonumber
\end{eqnarray}
Ignoring the multivalued nature of the logarithm and taking $\ln i = i \pi/2$, we obtain $\ln i x = \ln x + i \pi/2$ and hence
\begin{eqnarray} \label{Hur_dS}
\mathrm{Re} \mathcal{W}^{\rm (sp)}_{\rm dS} &=& - \mathrm{Im} \Bigl[ \zeta'(0, i \mu - i \kappa) + \zeta'(0, i \mu + i \kappa) - 2 \zeta'(0, 1/2 + i \mu) \Bigr]
\nonumber\\
&-& \frac{\pi}2 + (\mu - \kappa) \ln(\mu - \kappa) + (\mu + \kappa) \ln(\mu + \kappa) - 2\mu \ln(\mu),
\\
\mathrm{Im} \mathcal{W}^{\rm (sp)}_{\rm dS} &=& \mathrm{Re} \Bigl[ \zeta'(0, i \mu - i \kappa) + \zeta'(0, i \mu + i \kappa) - 2 \zeta'(0, 1/2 + i \mu) \Bigr]
\nonumber\\
&+& \frac12 \Bigl[ \ln(\mu - \kappa) + \ln(\mu + \kappa) \Bigr].
\end{eqnarray}
This representation is particularly useful for numerical evaluation of the one-loop effective action, as illustrated in Fig.~\ref{fig:ReImSpinordS2}.

Figure~\ref{fig:ReImSpinordS2} numerically shows the effect of spacetime curvature ($R=H^2$) and Maxwell scalar (${\cal F} = -F_{\mu \nu} F^{\mu \nu}/4 = E^2/2$) on the one-loop effective action and the vacuum persistence amplitude (vacuum decay). As shown in the left panel, the effect of curvature on the action is more significant than that of the Maxwell scalar. The effect of the Maxwell scalar is noticeable only when the curvature is small. The effect of curvature on vacuum decay via the vacuum persistence amplitude in the right panel is more significant than that of the Maxwell scalar, which implies that Gibbons-Hawking radiation decays the dS space more rapidly than the Schwinger effect except for small curvature and Maxwell scalar.
For large curvature, since the curvature dominates the Maxwell scalar, we may expand~\eqref{sp_vac_act} in a power series of $\kappa = q E/H^2$:
\begin{eqnarray}
\mathrm{Re} W^{\rm (sp)}_{\rm dS} = \mathrm{Re} W^{\rm (sp)}_{\rm dS} \Big|_{E=0} + \int_0^\infty \mathrm{e}^{-2 mx/H} \Bigl( \frac{\cos x}{\sin x} -\frac1{\sin x} \Bigr) dx \times \Bigl( \kappa^2 \frac{H}{m} \Bigr) + \cdots.
\end{eqnarray}
Similarly, we can also express~\eqref{Hur_dS} by expanding $\mu$ as $m/H + \kappa^2 H/2 m + \cdots$.

The Bogoliubov coefficients $\alpha$ for scalars and spinors in the global coordinates~\cite{Jiang:2020evx, Chen:2025xrv} contains an angular contribution that appears as a phase. In reconstructing the effective action from the vacuum persistence amplitude (i.e., from its imaginary part), as discussed in Sec.~\ref{sec3}, this phase generates an entire function in the effective action. Using the expression for $\alpha$ given in Ref.~\cite{Chen:2025xrv},
\begin{eqnarray} \label{alp-ds}
\alpha = 2^{i 4 \mu} \frac{4 \pi \mu}{\sqrt{\mu^2 - \kappa^2}}  \Bigl[ \frac{\Gamma(-2 i \mu)^2}{\Gamma(-i \kappa - i \mu) \Gamma(i \kappa - i \mu)} \Bigr] \times \Bigl[ \frac{1}{\Gamma(\ell + 1 - i \mu) \Gamma(- \ell - i \mu)} \Bigr],
\end{eqnarray}
the complex effective action, $W = - i \ln \alpha^*$, after discarding terms that are removed by renormalization, takes the form
\begin{eqnarray} \label{com-act-ds}
W &=& - i \Bigl[ 2 \ln \Gamma(2 i \mu) - \ln \Gamma(i \mu + i \kappa) - \ln \Gamma(i \mu - i \kappa)
\nonumber\\
&& - \ln \Gamma(\ell + 1 + i \mu) - \ln \Gamma(-\ell + i \mu) \Bigr].
\end{eqnarray}
Here, $\ell$ is an integer. Note that the last square bracket in~\eqref{alp-ds} has magnitude $\sinh (\pi \mu)/\pi$, independently of $\ell$, but its phase depends on $\ell$.\footnote{This result follows from the identity $\frac{1}{\Gamma(\ell + 1 - i \mu) \Gamma(- \ell - i \mu)} =  \mathrm{e}^{-i \pi \ell} \frac{\Gamma(\ell + 1 + i \mu)}{\Gamma(\ell + 1 - i \mu)} \times \frac{\sinh (\pi \mu)}{\pi}.$} Thus, the last two terms in~\eqref{com-act-ds} give a missing phase in Sec.~\ref{sec3}. To represent the effective action as the proper-time integral, we use the integral representation of the logarithm of the Gamma function~\cite{gradshteyn2014table} and analytically continue the integral to complex plane of $z$,
\begin{eqnarray}
\ln \Gamma(z) = \int_0^{\infty} \frac{ds}{s} \left[ \frac{\mathrm{e}^{-z s}}{1 - \mathrm{e}^{-s}} - \frac{\mathrm{e}^{-s}}{1 - \mathrm{e}^{-s}} + (z - 1) \, \mathrm{e}^{-s} \right].
\end{eqnarray}
Following the regularization procedure in Ref.~\cite{Kim:2008yt}, the first and last two terms in~\eqref{com-act-ds} contribute to
\begin{eqnarray} \label{sp-ent}
\mathrm{Re} \, W =  P \int_{0}^{\infty} \frac{ds}{s} \mathrm{e}^{- 2 \mu s} \left[ \frac{1}{\sin s} - \frac{1}{s} + \frac{\cos (s) - \cos ((2 \ell + 1) s)}{\sin s} \right].
\end{eqnarray}
The first two terms in the integral~\eqref{sp-ent} are the second principal value integral of the effective action~\eqref{sp_vac_act}, while the last term is regular at $s =0$ and does not contribute to the imaginary part, which corresponds to an entire function.

Actually, relevant quantum fluctuations are expected if one turns on large angular modes that are included as part of the regularized sum over $\ell$~\cite{Kim:2010cb}. These modes introduce inhomogeneities and subsequent uncontrolled interactions that may, in general, lead to non-manifestly invariant contributions. Remarkably, our results show that the part of the effective action associated with its pole structure (and hence with the vacuum persistence amplitude) is independent of angular modes. In this sense, we obtain the invariant and physically relevant part of the one-loop effective action for dS$_2$.

\section{Effective Actions of Spinor QED in Global AdS$_2$} \label{sec5}

In this section, we consider the Schwinger effect for the AdS$_2$ space in the global coordinates with the scalar curvature $R = - 2H^2$
\begin{eqnarray} \label{AdS2_metric}
ds^2 = - \frac{\cosh^2(H \rho)}{H^2} dt^2 + d\rho^2, \qquad 0 \le t \le 2 \pi, \quad 0 < \rho < \infty,
\end{eqnarray}
and a constant electric field with the gauge potential
\begin{eqnarray} \label{AdS2_A}
A = \frac{E \sinh(H \rho)}{H^2} dt \quad \Rightarrow \quad dA = - E \, \vartheta^t \wedge \vartheta^\rho,
\end{eqnarray}
where $\vartheta^t = \cosh (Ht) dt/H$ and $\vartheta^\rho = d \rho$.
Here, the electric field is chosen to point in the opposite direction to compare with dS spacetime, but the final result is independent of the direction of the field.

The Bogoliubov coefficients for spinor pair production in global AdS$_2$ spacetime determine both the mean number of produced pairs and the vacuum persistence amplitude, up to the spin multiplicity, as~\cite{Chen:2025xrv}
\begin{eqnarray}
{\cal N}^{\rm (sp)}_{\rm AdS} = |\beta|^2 = \Bigl( \frac{\cosh \pi \mu}{\cosh \pi \kappa} \Bigr)^2, \qquad |\alpha|^2 = \frac{\sinh(\pi \kappa - \pi \mu) \sinh(\pi \kappa + \pi \mu)}{\cosh^2\pi \kappa}.
\end{eqnarray}
Here, the parameters $\kappa$ and $\mu$ are defined by
\begin{equation}
\kappa = q E/H^2, \qquad \mu = \sqrt{\kappa^2 - m^2/H^2} < \kappa.
\end{equation}
We have assumed a violation of the Breitenlohner–Freedman (BF) bound, $\kappa > m/H$, under which Schwinger pair production occurs~\cite{Pioline:2005pf}.

Notably, the Bogoliubov coefficients for spinor QED in AdS$_2$ can be obtained from those in dS$_2$ by simple interchange $\kappa \leftrightarrow \mu$. Consequently, the corresponding complex one-loop QED effective action takes the form
\begin{eqnarray} \label{cW_AdS}
W^{\rm (sp)}_{\rm AdS} &=& \frac{i}{2} \int_0^\infty \left( \mathrm{e}^{- 2 i (\kappa- \mu) y} + \mathrm{e}^{- 2 i (\kappa + \mu) y} \right) \left( \frac{\cosh y}{\sinh y} - \frac1{y} \right) \frac{dy}{y}
\nonumber\\
&-& i \int_0^\infty \mathrm{e}^{- 2 i \kappa y} \left( \frac1{\sinh y} - \frac1{y} \right) \frac{dy}{y},
\end{eqnarray}
and its real part is
\begin{eqnarray} \label{rW_AdS}
\mathrm{Re} W^{\rm (sp)}_{\rm AdS} &=& - \frac12 P \int_0^\infty \left( \mathrm{e}^{-2 (\kappa - \mu) x} + \mathrm{e}^{-2 (\kappa + \mu) x} \right) \left( \frac{\cos x}{\sin x} - \frac1{x} \right) \frac{dx}{x}
\nonumber\\
&+& P \int_0^\infty \mathrm{e}^{-2 \kappa x} \left( \frac1{\sin x} - \frac1{x} \right) \frac{dx}{x}.
\end{eqnarray}

It is straightforward to see that the complex one-loop effective action in AdS, Eq.~\eqref{cW_AdS}, reduces to its Minkowski counterpart, Eq.~\eqref{cW_Min}, in the limit $H \to 0$. By contrast, the pure AdS result cannot be obtained simply by taking the limit $\kappa \to 0 \; (E = 0)$. This is because both the mean particle number and the vacuum persistence amplitude were derived under the assumption that $\mu$ is real, which requires $\kappa > m/H$.

A power-law expansion of $\mathrm{Re} W^{\rm (sp)}_{\rm AdS}$ can be obtained in the regime where the exponential terms decay sufficiently rapidly. This expansion follows straightforwardly from Eq.~\eqref{sp_exp} upon interchanging $\kappa$ and $\mu$
\begin{eqnarray}
\mathrm{Re} W^{\rm (sp)}_{\rm AdS} &=& \frac{3 \kappa^2 - \mu^2}{12 \kappa (\kappa^2 - \mu^2)} + \frac{15 \kappa^6 + 3 \mu^2 \kappa^4 + 21 \mu^4 \kappa^2 - 7 \mu^6}{1440 \kappa^3 (\kappa^2 - \mu^2)^3}
\nonumber\\
&+& \frac{63 \kappa^{10} + 165 \mu^2 \kappa^8 + 470 \mu^4 \kappa^6 - 310 \mu^6 \kappa^4 + 155 \mu^8 \kappa^2 - 31 \mu^{10}}{20160 \kappa^5 (\kappa^2 - \mu^2)^5} + \cdots.
\end{eqnarray}
Two special limits are of particular interest. The first is the weak-curvature limit $H \to 0$, in which
\begin{equation}
\mathrm{Re} W^{\rm (sp)}_{\rm AdS} \Big|_{H \to 0} = \frac{q E}{6 m^2} + \frac{q^3 E^3}{45 m^6} + \frac{8 q^5 E^5}{315 m^{10}} + \left( \frac1{12 q E} - \frac{q E}{60 m^4} - \frac{2 q^3 E^3}{63 m^8} \right) H^2 + \mathcal{O}(H^4).
\end{equation}
By contrast, the weak-field limit $E \to 0$ is unphysical: it renders $\mu$ imaginary due to the Breitenlohner–Freedman (BF) bound. Consequently, pair production does not occur in this limit.

\begin{figure}[t]
\centering
\setlength{\fboxrule}{1.2pt} 
\setlength{\fboxsep}{6pt} 

\begin{minipage}{0.48\textwidth}
\centering
\fbox{%
\begin{minipage}{\linewidth}
\centering
\includegraphics[width=\linewidth]{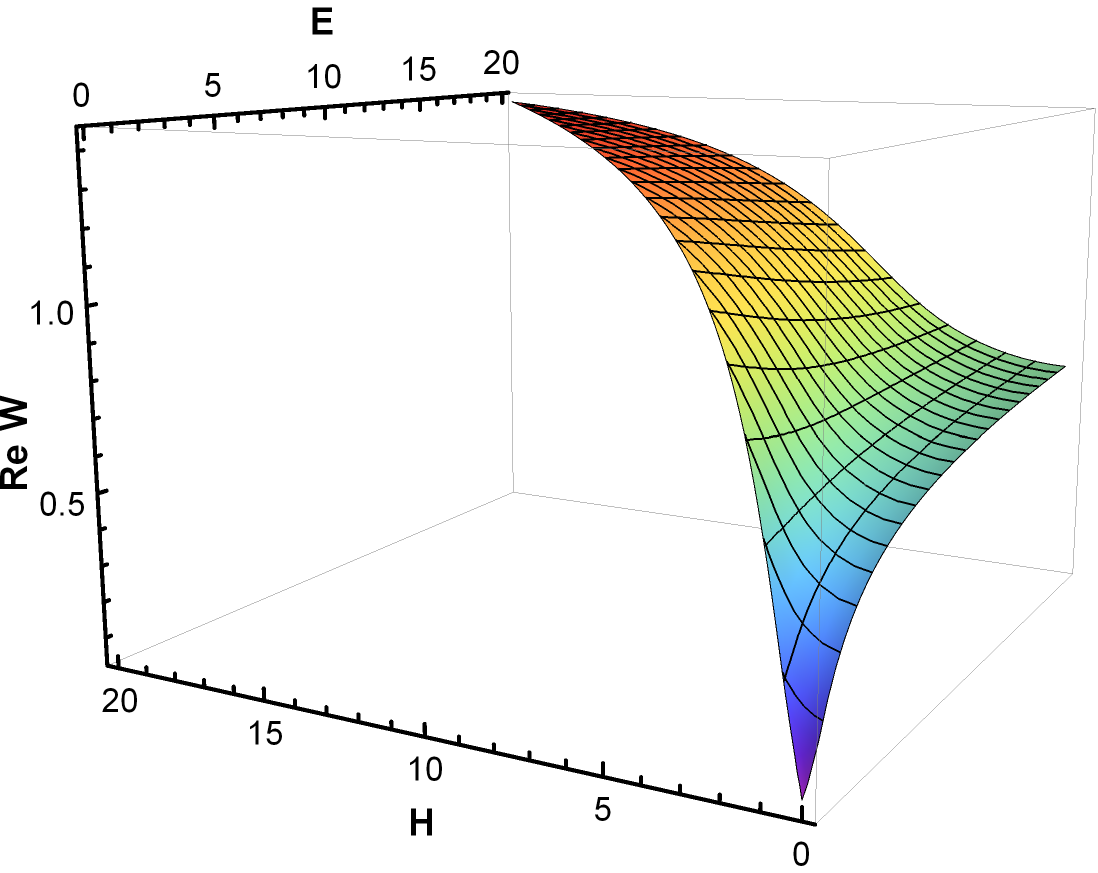}
\vspace{2mm}\textbf{(a)}\par
\end{minipage}}
\end{minipage}
\hfill
\begin{minipage}{0.48\textwidth}
\centering
\fbox{%
\begin{minipage}{\linewidth}
\centering
\includegraphics[width=\linewidth]{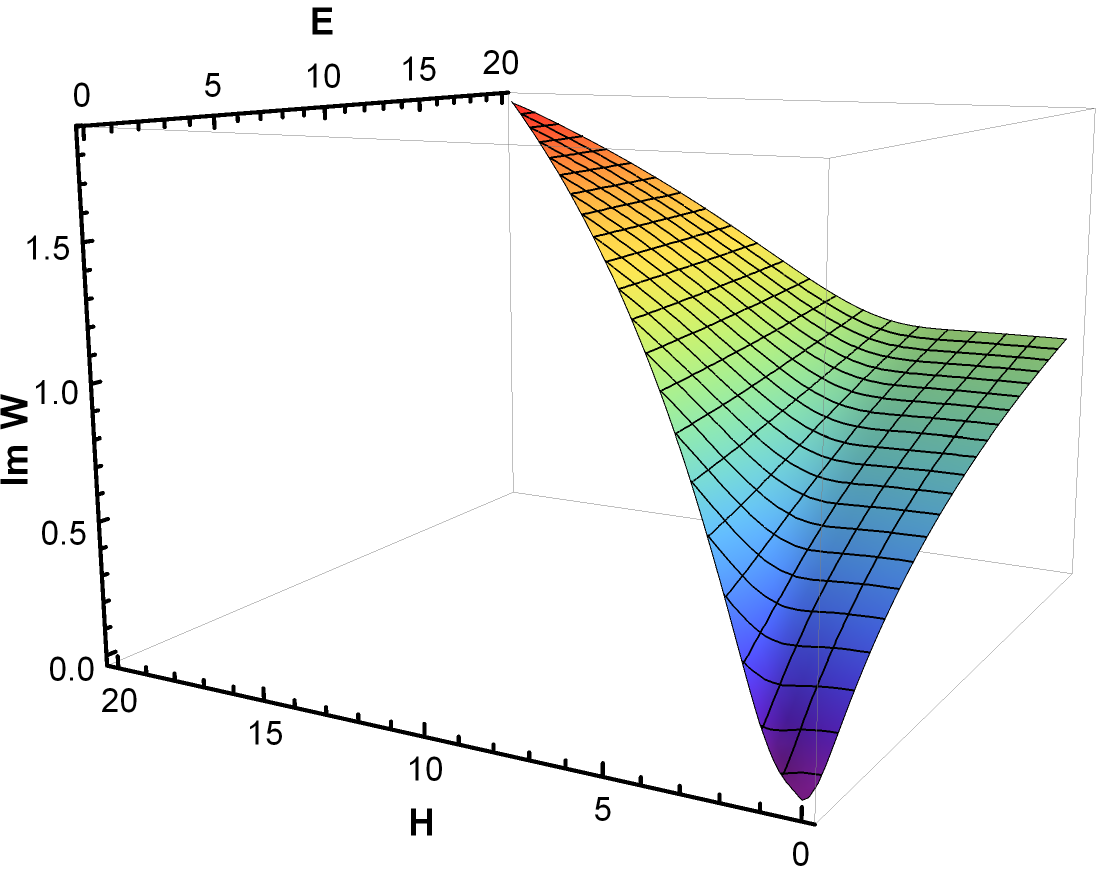}
\vspace{2mm}\textbf{(b)}\par
\end{minipage}}
\end{minipage}

\caption{The plots of real part (a) and imaginary part (b) of $\mathcal{W}^{\rm (sp)}_{\rm AdS}$ against $E$ and $H$ for spinor QED in AdS$_2$ in unit of $m = 1$ and $q = 1$. The necessary condition for pair production is $q E > m H$, violation of BF bound. [Left panel] The effective action increases more significantly as $H$ increases than as $E$ increases. Except for the small curvature region and Maxwell scalar, the curvature effect dominates over the effect of the Maxwell scalar in the region that violates the BF bound. [Right panel] The vacuum decays more rapidly for large curvature than for large Maxwell scalar.}
\label{fig:ReImSpinorAdS2}
\end{figure}

An alternative representation of the real and imaginary parts of the one-loop effective action, expressed in terms of the Hurwitz zeta function, is given by
\begin{eqnarray} \label{Hur_AdS}
\mathrm{Re} \mathcal{W}^{\rm (sp)}_{\rm AdS} &=& - \mathrm{Im} \Bigl[ \zeta'(0, i \kappa - i \mu) + \zeta'(0, i \kappa + i \mu) - 2 \zeta'(0, 1/2 + i \kappa) \Bigr]
\nonumber\\
&-& \frac{\pi}2 + (\kappa - \mu) \ln(\kappa - \mu) + (\kappa + \mu) \ln(\kappa + \mu) - 2 \kappa \ln(\kappa),
\\
\mathrm{Im} \mathcal{W}^{\rm (sp)}_{\rm AdS} &=& \mathrm{Re} \Bigl[ \zeta'(0, i \kappa - i \mu) + \zeta'(0, i \kappa + i \mu) - 2 \zeta'(0, 1/2 + i \kappa) \Bigr]
\nonumber\\
&+& \frac12 \Bigl[ \ln(\kappa - \mu) + \ln(\kappa + \mu) \Bigr].
\end{eqnarray}
Using this representation, the one-loop effective action is illustrated in Fig.~\ref{fig:ReImSpinorAdS2}.

The effect of spacetime curvature and Maxwell scalar on QED action in AdS space is shown in Fig.~\ref{fig:ReImSpinorAdS2}. The essential difference from the dS space is the existence of the Breitenlohner-Freedman (BF) bound, below which the stability of the AdS space is guaranteed in the presence of an electric field. Thus, the region for the BF bound does not produce any pairs and is thus excluded from the numerical data. Similarly to the dS space, the QED action increases more significantly as the curvature increases than as the Maxwell scalar increases. The curvature effect is also more significant than the Maxwell scalar effect.

One may raise the intriguing question: can one find effective action in a pure AdS space without an electric field probed by the spinor field? The pure AdS space belongs to the BF bound and does not produce pairs. This is analogous to QED actions in a Minkowski spacetime either in a constant magnetic field or a constant electric field. A constant electric field produces Schwinger pairs, so the QED action can be obtained from the vacuum persistence amplitude. However, a constant magnetic field does not produce pairs, and the magnetic vacuum is stable at the one-loop level, so the the inverse procedure of obtaining the QED action from the the vacuum persistence amplitude cannot be used. Instead, the Jost function may be used to find the inverse scattering matrix for magnetic fields in Euclidean time, and the in-out formalism applies to the scattering matrix~\cite{Kim:2011cx}. Note that QED actions still have duality under $E = -i B$. Similarly to electromagnetic duality in QED actions, there is another duality between $H$ in dS space and $i H$ in the AdS space, and the effective action would take the form of
\begin{eqnarray} \label{eff_pureAdS}
W^{\rm (sp)}_{\rm AdS} \Big|_{E=0} = -i W^{\rm (sp)}_{\rm dS} \Big|_{E=0}\left( H \to i H \right) = \int_0^\infty \mathrm{e}^{- 2 m y/H} \left( \frac{\cosh y}{\sinh y} - \frac{1}{\sinh y} \right) \frac{dy}{y}.
\end{eqnarray}
Note that the effective action \eqref{eff_pureAdS} takes the form of a difference between the spinor and scalar QED actions in a constant magnetic field~\cite{Kim:2011cx}, in the correspondence $H/2 \leftrightarrow qB/m$.

\section{Connection of QED actions between dS$_2$ and AdS$_2$} \label{sec6}
One of the essential differences between the dS and AdS space is the stability probed by a field: the dS space produces particle pairs with Gibbons-Hawking temperature, whereas the AdS space is stable against any particle production. Recently, it has been shown that the Schwinger effect by an electric field connects these two topologically different spaces. The mean numbers of spinors or scalars produced satisfy the reciprocal relation, according to which the mean number in the AdS space is the inverse of that in the dS space or vice versa, provided that the spacetime curvature $R$ is analytically continued~\cite{Chen:2025xrv},
\begin{eqnarray}\label{rec-rel}
{\cal N}^{\rm (sp)}_{\rm dS} (E, R) {\cal N}^{\rm (sp)}_{\rm AdS} (E, R) = 1.
\end{eqnarray}
The same relation holds for scalars. From now on, we will focus on the spinor case.

The reciprocal relation is rooted in the boundary conditions: scattering of a probe field over the barrier in the dS space and tunneling through the barrier in the AdS space. In the dS space, a positive frequency incoming mode with amplitude $C$ from the past is scattered by a barrier of dS curvature and a uniform electric field and splits into a positive  frequency outgoing mode with amplitude $A$ and a negative frequency outgoing mode with amplitude $B$ in the future.
Then, the Bogoliubov coefficients are
\begin{eqnarray}
|\alpha_{\rm dS}|^2 = \frac{|A|^2}{|C|^2}, \qquad |\beta_{\rm dS}|^2 = \frac{|B|^2}{|C|^2}.
\end{eqnarray}
On the other hand, in the AdS space, the tunneling boundary condition for spinors with respect to group velocity and taking into account the Klein paradox prescribes the following flux~\cite{Kim:2003qp}
\begin{eqnarray}
|B|^2 = |A|^2 + |C|^2,
\end{eqnarray}
and therefrom
\begin{eqnarray}
|\alpha_{\rm AdS}|^2 = \frac{|A|^2}{|B|^2}, \qquad |\beta_{\rm AdS}|^2 = \frac{|C|^2}{|B|^2}.
\end{eqnarray}
Therefore, the reciprocal relation~\eqref{rec-rel} holds for spinors, and it holds similarly for scalars.

Then, one may raise the question whether the vacuum persistence amplitudes and thereby the effective actions are connected between the dS and AdS spaces. We may apply the reciprocal relation to the vacuum persistence amplitude for the AdS space and express it in terms of those quantities in dS space
\begin{eqnarray}
2 \mathrm{Im} W^{\rm (sp)}_{\rm AdS} = - \ln (1 - |\beta_{\rm AdS}|^2)
= - \ln \Bigl(- \frac{|\alpha_{\rm dS}|^2}{|\beta_{\rm dS}|^2} \Bigr).
\end{eqnarray}
Here, the minus sign requires that $\mu$ and $\kappa$ interchange their roles: $\mu \leftrightarrow \kappa$, as expected in Eqs.~\eqref{spinor_ds} and \eqref{spinor_ds_alp}. Thus, the QED action in the AdS space can be obtained by introducing $\mu \leftrightarrow \kappa$ into the QED action in the dS space.

\begin{figure}
    \centering
    \includegraphics[scale=0.8, angle=0]{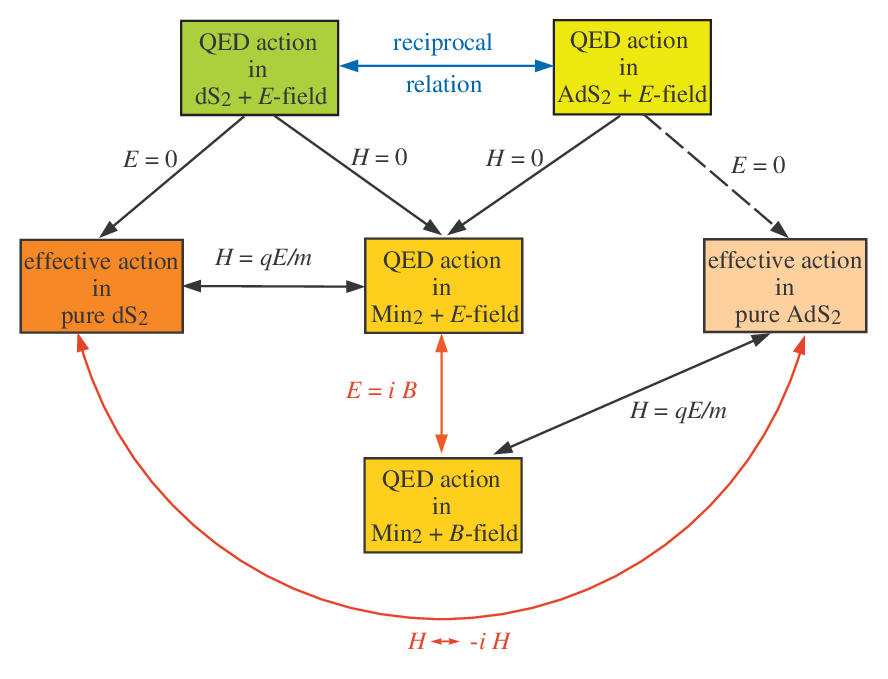}
    \caption{Connection of QED actions between dS$_2$ and AdS$_2$ and reduction to QED action in Minkowski spacetime and pure dS$_2$.}
    \label{fig:diagram}
\end{figure}

We may thus argue that QED actions connect two topologically different spaces, dS$_2$ and AdS$_2$, as summarized in Fig.~\ref{fig:diagram}. The vacuum persistence amplitude is related by the reciprocal relation under analytical continuation of the curvature. The effective action of the pure dS space, which is the limit of $E = 0$, has the same form as the QED action in a constant electric field in two-dimensional Minkowski space under the correspondence $H/m \leftrightarrow qE/m^2$. The effective action of pure AdS space, which cannot be obtained by the limit of $E = 0$ contrary to the dS space, has the same form as QED action in a constant magnetic field that can be obtained by reduction from four dimensions or analytical continuation $E \leftrightarrow iB$ in two dimensions, and does not have the vacuum persistence amplitude (twice the imaginary part). As expected, the effective actions in dS space and AdS space are an analytical continuation under $H \leftrightarrow -iH $.

\begin{figure}
\begin{minipage}{0.48\textwidth}
\centering
\fbox{%
\begin{minipage}{\linewidth}
\centering
\includegraphics[width=\linewidth]{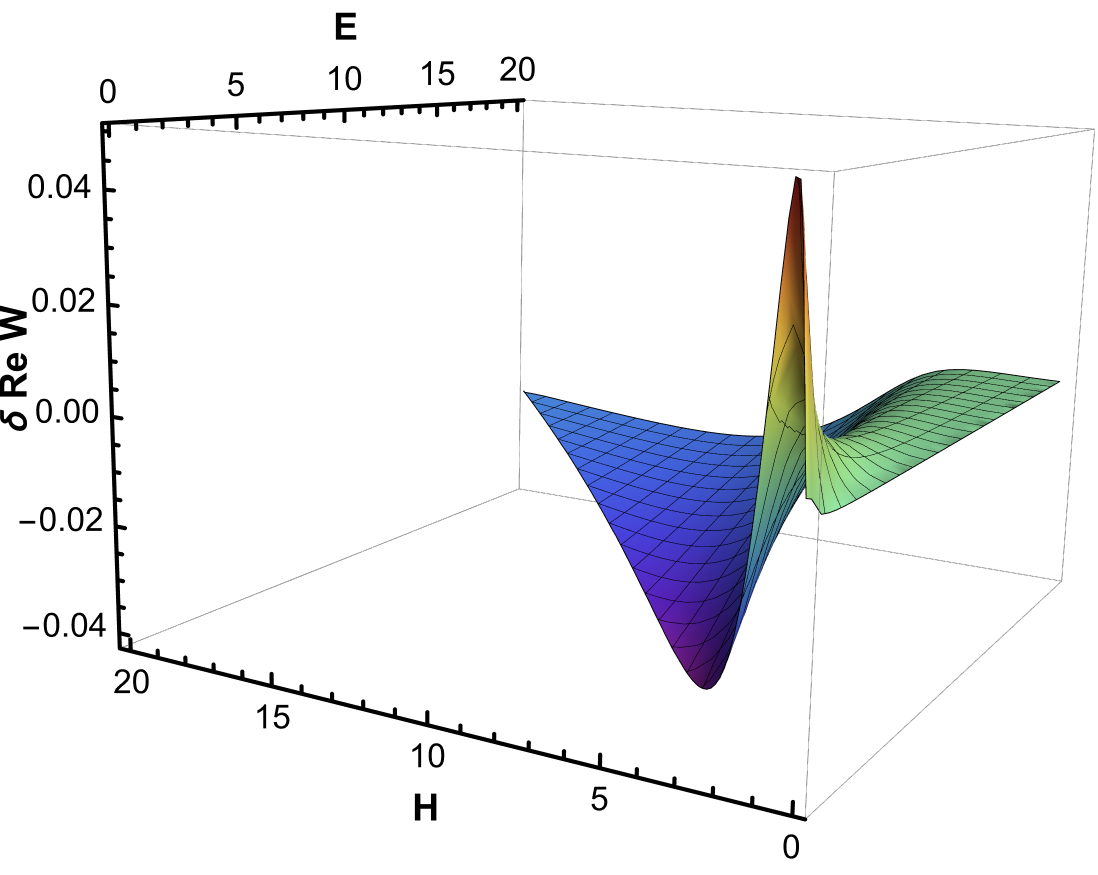}
\vspace{2mm}\textbf{(a)}\par
\end{minipage}}
\end{minipage}
\hfill
\begin{minipage}{0.48\textwidth}
\centering
\fbox{%
\begin{minipage}{\linewidth}
\centering
\includegraphics[width=\linewidth]{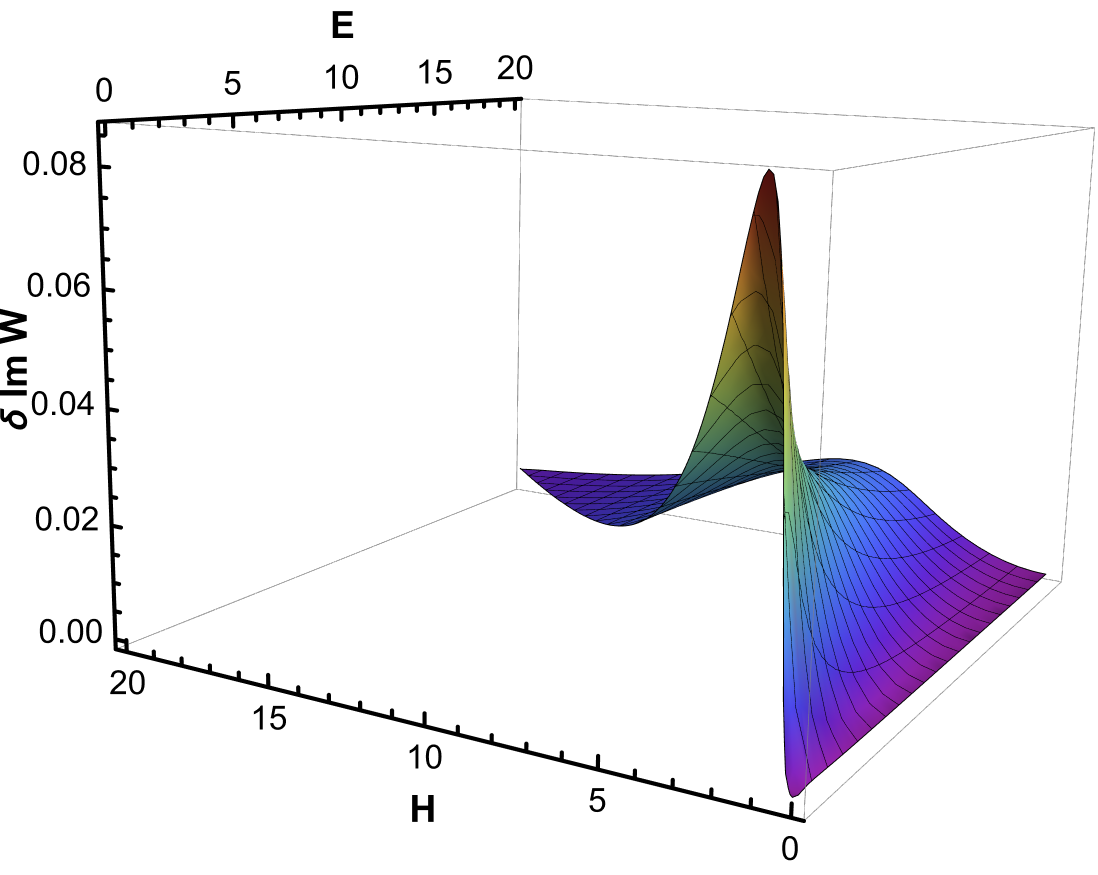}
\vspace{2mm}\textbf{(b)}\par
\end{minipage}}
\end{minipage}

\caption{The plots show the difference between real parts (a) [left panel] and imaginary parts (b) [right panel] of $\mathcal{W}^{\rm (sp)}_{\rm dS}$ and $\mathcal{W}^{\rm (sp)}_{\rm AdS}$ against $E$ and $H$ for spinor QED in dS$_2$ and AdS$_2$ in units of $m = 1$ and $q = 1$. In both plots the domains are constrained by the necessary condition for pair production in AdS$_2$  that is $q E > m H$, violation of BF bound.}
\label{fig:DiffReImSpinordS2AdS2}
\end{figure}

The difference of QED actions and the vacuum persistence amplitudes between the dS and AdS spaces is shown in Fig.~\ref{fig:DiffReImSpinordS2AdS2}, where the region for the BF bound is excluded. In the region of small curvatures, where the Schwinger effect dominates the Gibbons-Hawking radiation, the QED actions are close to each other, whereas the QED action in the AdS space is larger than that in the dS space as the curvature and Maxwell scalar increase. On the other hand, the vacuum persistence amplitude is always larger in the dS space than in the AdS space, and the difference is more significant near the curve of the BF bound than in the region of large curvature and Maxwell scalar.

\section{Conclusion} \label{sec7}
In the in-out formalism the exact one-loop effective action can be found from scattering matrix between the out-vacuum and in-vacuum in properly defined regions, which is formally expressed in terms of the Bogoliubov coefficient. Instead, we have introduced a novel method that uses the correspondence between the vacuum persistence amplitude (twice the imaginary part of effective action) and the real part of effective action. This method directly gives a proper-time integral representation of effective action introduced by Schwinger, modulo some entire function, once the exact mean number of spontaneously produced pairs by background fields of spacetime curvature and electric field is determined, which is a kind of inverse scattering procedure.

We have found the exact one-loop QED actions probed by a charged scalar or spinor field in a uniform electric field in the global (A)dS$_2$. Two dimensionless parameters determine the relativistic instanton actions from the Klein-Gordon or Dirac equations: one is the ratio of the Maxwell scalar to the spacetime curvature square and the other is the ratio of the mass square to the spacetime curvature. In terms of these parameters, QED actions are expressed both in the proper-time integrals~\eqref{sp_vac_act},~\eqref{rW_AdS} and in the Hurwitz zeta functions~\eqref{Hur_dS},~\eqref{Hur_AdS}. In the zero-curvature limit, the QED actions in dS$_2$ and AdS$_2$ recover that in the two-dimensional Minkowski space. That QED actions depend on two characteristic parameters responsible for the mean numbers of produced pairs exhibits a strong interplay between gravity and Maxwell field, which cannot be realized in perturbative expansion methods.

In the zero-field limit the QED action in dS$_2$ reduces to the effective action of pure dS$_2$. Interestingly, the one-loop effective action~\eqref{eff_pure_dS} or Lagrangian (density) for pure dS$_2$ reduced from the QED action has the form of the difference between spinor and scalar QED actions in a constant electric field in a Minkowski space, in the correspondence $H/2 \leftrightarrow qE/m$, which results in the mean number of Gibbons-Hawking radiation in the global coordinates of dS space.
Similarly to duality $E \leftrightarrow -iB$ in QED actions, there is duality of effective actions between the dS space and AdS space in the correspondence $H \leftrightarrow iH$. In fact, the effective action in pure AdS space takes the form of difference between the spinor and scalar QED actions in a constant magnetic field in a Minkowski spacetime, in the correspondence $H/2 \leftrightarrow qB/m$.
This analogy explains the reason why the dS space is unstable against pair production, while the AdS space is stable, confirming the Breitenlohner-Freedman (BF) bound.

We have shown that the reciprocal relations hold between the dS$_2$ and AdS$_2$ spaces, not only for the mean numbers, but also for the vacuum persistence amplitudes. These are a consequence of boundary conditions imposed on probing spinor fields between the in-region and out-region: scattering over a barrier from spacetime curvature and electric field in dS space, whereas tunneling under the barrier in AdS space. The reciprocal relation explains why the QED actions in dS and AdS spaces have the same form under the interchange of parameters for instanton actions. All of these imply that QED actions and the mean numbers in dS$_2$ and AdS$_2$ spaces are connected, as shown in Fig.~\ref{fig:diagram}.

Finally, the present framework suggests that the invariant content of one-loop effective actions in maximally symmetric two-dimensional spacetimes can be extracted directly from the vacuum persistence amplitude, without relying on detailed mode-by-mode constructions. This perspective may be useful for studying quantum effects in more general backgrounds, including effective descriptions of the near-horizon geometries of (near-)extremal charged black holes using holographic methods~\cite{Heydeman:2020hhw}, where strong gauge fields and spacetime curvature coexist.

\appendix

\acknowledgments

C.M.C. and S.P.K. thank Chul Min Kim for helpful discussions, the participants of the joint program [APCTP-2025-J01] held at APCTP, Korea for fruitful discussions, and Ehsan Bavarsad for sharing a manuscript on scalar QED action in four-dimensional de Sitter space. C.M.C. thanks Kyung Taec Kim for the warm hospitality at CoReLS-IBS, Korea.
The work of C.M.C. was supported by the National Science and Technology Council of the R.O.C. (Taiwan) under the grant NSTC 114-2112-M-008-010. The work of S.P.K. was supported by the Institute of Basic Science (Grant No. IBSR038-D1).


\end{document}